# An analytic relation for the thickness of accretion flows

CAI ZhenYi, GU WeiMin[†], XUE Li & LU JuFu

Department of Physics and Institute of Theoretical Physics and Astrophysics, Xiamen University, Xiamen 361005, China

**We take the vertical distribution of the radial and azimuthal velocity into account in spherical coordinates, and find that the analytic relation $c_{s0}/v_K\Theta = [(\gamma-1)/2\gamma]^{1/2}$ is valid for both geometrically thin and thick accretion flows, where $c_{s0}$ is the sound speed on the equatorial plane, $v_K$ is the Keplerian velocity, $\Theta$ is the half-opening angle of the flow, and $\gamma$ is the adiabatic index.**

accretion disks, black hole physics, hydrodynamics

Accretion of rotating matter onto a compact object powers many energetic astrophysical systems, such as X-ray binaries, active galactic nuclei (AGNs), and ultraluminous X-ray source (ULXs). The standard thin disk model[1] has been successfully applied to cataclysmic variables[2-3]. The other two well-known accretion models are the radiation-supported, optically thick advection-dominated accretion flows (ADAFs) or slim disks[4] and the ion-supported, optically thin ADAFs[5] with considering the mass accretion and the energy advection. The optically thick type of ADAFs has been successfully applied to narrow-line Seyfert 1 galaxies[6] and ULXs[7]; while the optically thin type of ADAFs has been widely used to black hole X-ray binaries[8] and low-luminosity AGNs[9]. These two types of ADAFs are both regarded to be slim, i.e., intermediate between thin and thick with $H/R \sim 1$, where $H$ is the half-thickness of the flow, and $R$ is the cylindrical radius.

In the previous work of Gu & Lu[10] (hereafter GL07), in cylindrical coordinates $(R, z, \varphi)$, we discussed the potential importance of taking the explicit form of the gravitational potential for calculating slim disk solutions. As we discussed in GL07, we pointed out that the Hōshi form of the potential[11],

$$\psi(R,z) \cong \psi(R,0) + \frac{1}{2}\Omega_K^2 z^2, \qquad (1)$$

is valid only for geometrically thin disks with $H/R < 0.2$. Thus the well-known relationship $c_{s0}/\Omega_K H = \text{constant}$[3] does not hold for slim disks with $H/R \sim 1$, where $c_{s0}$ is the sound speed on the equatorial plane, and $\Omega_K = (GM/r^3)^{1/2}$ is the Keplerian angular velocity with $M$ being the mass of the central black hole. We also showed that, the slim disk solutions under the explicit form are geometrically much thicker than the classical slim disk solutions under the Hōshi form. In particular, the thickness may go to infinite thus no solution exists. Such a result is not difficult to understand since the Hōshi form greatly magnifies the vertical gravitational force, thus the classical slim disks are restricted to be geometrically slim with $H/R \sim 1$. Except for the form of the potential, GL07 kept exactly the same method as Kato et al.[3] for calculations. The method is, however, based on the following simple vertical hydrostatic equilibrium:

$$\frac{1}{\rho}\frac{\partial p}{\partial z} + \frac{\partial \psi}{\partial z} = 0, \qquad (2)$$

instead of the general form of the vertical momentum equation[14]:

$$\frac{1}{\rho}\frac{\partial p}{\partial z} + \frac{\partial \psi}{\partial z} + v_R\frac{\partial v_z}{\partial R} + v_z\frac{\partial v_z}{\partial z} = 0, \qquad (3)$$

where $\rho$ is the mass density, $p$ is the pressure, $v_R$ is the cylindrical radial velocity and $v_z$ is the vertical velocity.

We argue that equation (2) is valid only for geometrically thin disks because in this case there exists the

Received; accepted
doi:
[†]Corresponding author (email: guwm@xmu.edu.cn)
Supported by the National Basic Research Program of China (Grant No. 2009CB824800) and the National Natural Science Foundation of China (Grant Nos. 10778711 and 10833002) relation $v_z \ll v_R$, thus $v_z$ can be ignored and equation (3) is simplified to equation (2). For geometrically slim flows, equation (2) no longer works well due to the following reason. As pointed out by Abramowicz et al.[12], there should be no component of the total velocity orthogonal to the surface, i.e., at the surface there exists the condition: $v_z/v_R = dH/dR$. Moreover, the stationary accretion flows calculated in realistic two- and three-dimensional simulations resemble quasi-spherical flows ($H/R \approx$ constant) much more than quasi-horizontal flows ($H \approx$ constant). Therefore, we have the relation $v_z/v_R \sim H/R$, which implies that, at the surface, $v_z$ is comparable to $v_R$ for geometrically slim flows and should not be ignored.

We revisit the vertical structure of thick accretion flows under the general form of the momentum equation. However, we are working on the general momentum equations in spherical coordinates instead of the equation (3) in cylindrical coordinates. Of course, the final results do not depend on the the coordinates system used. The reason why we adopt the spherical coordinates is that our work is based on the same self-similar assumptions as Narayan & Yi[13], which imply that the flow is quasi-spherical. Furthermore, a consequent improvement is that we take the vertical distribution of the radial and azimuthal velocity into account.

## 1 Equations

We consider a steady state axisymmetric accretion flow in spherical coordinates ($r$, $\theta$, $\varphi$), i.e., $\partial/\partial t = \partial/\partial \varphi = 0$, with $\theta = \pi/2$ corresponding to the equatorial plane. Here we adopt the Newtonian potential $\psi = -GM/r$ since it is convenient for the self-similar assumptions. The momentum equations are the following[14]:

$$v_r \frac{\partial v_r}{\partial r} + \frac{v_\theta}{r}(\frac{\partial v_r}{\partial \theta} - v_\theta) - \frac{v_\varphi^2}{r} = -\frac{GM}{r^2} - \frac{1}{\rho}\frac{\partial p}{\partial r}, \quad (4)$$

$$v_r \frac{\partial v_\theta}{\partial r} + \frac{v_\theta}{r}(\frac{\partial v_\theta}{\partial \theta} + v_r) - \frac{v_\varphi^2}{r}\cot\theta = -\frac{1}{\rho r}\frac{\partial p}{\partial \theta}, \quad (5)$$

$$v_r \frac{\partial v_\varphi}{\partial r} + \frac{v_\theta}{r}\frac{\partial v_\varphi}{\partial \theta} + \frac{v_\varphi}{r}(v_r + v_\theta \cot\theta) = \frac{1}{\rho r^3}\frac{\partial}{\partial r}(r^3 T_{r\varphi}), \quad (6)$$

where $v_r$, $v_\theta$, $v_\varphi$ are the three components of the velocity. Here, we only consider the $r\varphi$-component of the viscous stress tensor, $T_{r\varphi} = \rho v r \partial(v_\varphi/r)/\partial r$, where $v = \alpha c_s^2 r/v_K$ is the kinematic coefficient of viscosity[13], with $\alpha$ being the Shakura-Sunyaev viscosity parameter, $c_s = (p/\rho)^{1/2}$ being the sound speed, and $v_K = (GM/r)^{1/2}$ being the Keplerian velocity.

Due to mathematical complication, it is difficult to directly solve the above two-dimensional equations. To our knowledge, there are two methods to deal with this problem. One is to do analyses by employing the Taylor expansion[15], which, apparently, is valid only for geometrically thin disks, and the other is to solve the equations in the vertical (or '$\theta$') direction by taking the self-similar assumptions in the radial direction[13]. Our work follows the latter since the objects of study are definitely not geometrically thin. In the present paper, we do not take outflow into consideration and assume a hydrostatic equilibrium in '$\theta$' direction, i.e., $v_\theta = 0$, which agrees with the continuity equation. The self-similar assumptions are the following[13]:

$$v_r \propto r^{-1/2}; v_\varphi \propto r^{-1/2};$$
$$\rho \propto r^{-3/2}; c_s \propto r^{-1/2}.$$

With the above assumptions, equations (4-6) can be simplified as follows:

$$\frac{1}{2}v_r^2 + \frac{5}{2}c_s^2 + v_\varphi^2 - v_K^2 = 0, \quad (7)$$

$$\frac{1}{\rho}\frac{dp}{d\theta} = v_\varphi^2 \cot\theta, \quad (8)$$

$$v_r = -\frac{3}{2}\frac{\alpha c_s^2}{v_K}. \quad (9)$$

Comparing the two vertical hydrostatic equilibrium, in our opinion, equation (8) is better than equation (2) because the former is simplified from the general form by $v_\theta = 0$, whereas the latter is based on $v_z = 0$.

Different from Nayaran & Yi[13], we do not assume in advance a unified strength of energy advection. Instead, we follow Gu & Lu[10] and vary the geometrical thickness of the flow, i.e., the half-opening angle $\Theta$, which is $\Theta = \pi/2 - \theta$ (similar to varying $H/R$ in Gu & Lu[10]), to see the corresponding variation of the strength of energy advection. As many previous works[3,10], we assume a polytropic relation of state in the vertical direction: $p = K\rho^\gamma$, where $K$ is a constant and $\gamma$ is the adiabatic index. Then the equation (8) is transferred as:

$$\frac{dc_s^2}{d\theta} = \frac{\gamma - 1}{\gamma}v_\varphi^2 \cot\theta. \quad (10)$$

In addition, a boundary condition is required for solving the differential equation (10), which is set to be $c_s = 0$ (accordingly $\rho = 0$ and $p = 0$)[3] at the surface of the flow. After integrating equation (10) of $\theta$ with limits $c_s(\pi/2-\Theta) = 0$ and $c_s(\pi/2) = c_{s0}$, where $c_{s0} = (p_0/\rho_0)^{1/2}$ with the subscript 0 represents quantities on the equatorial plane, we obtain an analytic relationship:

$$\left(\frac{c_{s0}}{\upsilon_K \Theta}\right)^2 = \frac{f(\Theta)-1}{(\beta-5/4)+(\beta+5/4)f(\Theta)}\frac{1}{\Theta^2}, \quad (11)$$

where $f(\Theta) = \exp[-2\beta\frac{\gamma-1}{\gamma}\ln(\cos\Theta)]$ and $\beta = \sqrt{\frac{9\alpha^2}{8}+\frac{25}{16}}$.

The above equation is accurate but complicated. For small $\Theta$, we get a simplified relation by using Taylor expansion to equation (11):

$$c_{s0}/\upsilon_K\Theta = [(\gamma-1)/2\gamma]^{1/2}. \quad (12)$$

For geometrically thin disks, the above equation is identical with the relation $c_{s0}/\Omega_K H = [(\gamma-1)/2\gamma]^{1/2}$ in cylindrical coordinates[3].

## 2 Results and discussion

Our numerical procedure is that by varying $\Theta$ continuously, we will see the relationship between $\Theta$ and $c_{s0}/\upsilon_K\Theta$ for different $\gamma$. Moreover, to evaluate the validity of the analytic relation of equation (12), we show the variation of $c_{s0}/\upsilon_K\Theta$ with $\gamma$ for different half-opening angles $\Theta$. In our calculations, the viscosity parameter is fixed as $\alpha = 0.1$.

Figure 1 shows the variation of $c_{s0}/\upsilon_K\Theta$ with $\Theta$ for $\gamma = 4/3$ (solid line), $\gamma = 3/2$ (dashed line), and $\gamma = 5/3$ (dotted line). From the figure, we can see that the value of $c_{s0}/\upsilon_K\Theta$ does not change much for different half-opening angles, i.e., $c_{s0}/\upsilon_K\Theta \approx$ constant. Since we have obtained equation (12), which should be accurate for small $\Theta$, thus we can expect that the analytic relation of equation (12) is valid for both geometrically thin and thick flows. On the contrary, the relation $c_{s0}/\Omega_K H =$ constant obtained by using equation (1) in cylindrical coordinates[3] does not hold for the slim disks with $H/R \sim 1$ as long as taking the explicit form of the gravitational potential instead of the Hōshi form of the potential[10].

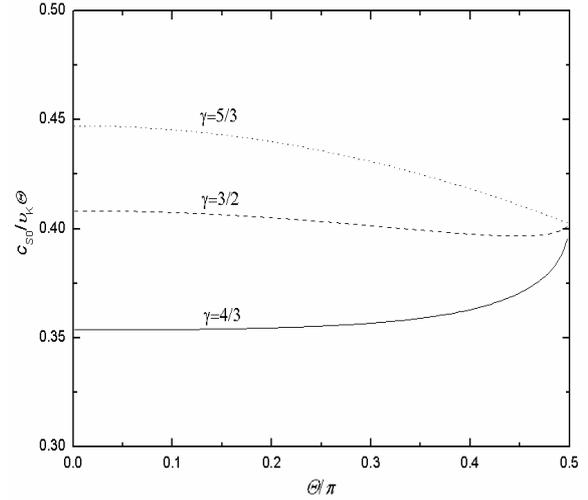

**Figure 1** Variation of $c_{s0}/\upsilon_K\Theta$ with $\Theta$ for $\gamma = 4/3$ (solid line), $\gamma = 3/2$ (dashed line), and $\gamma = 5/3$ (dotted line).

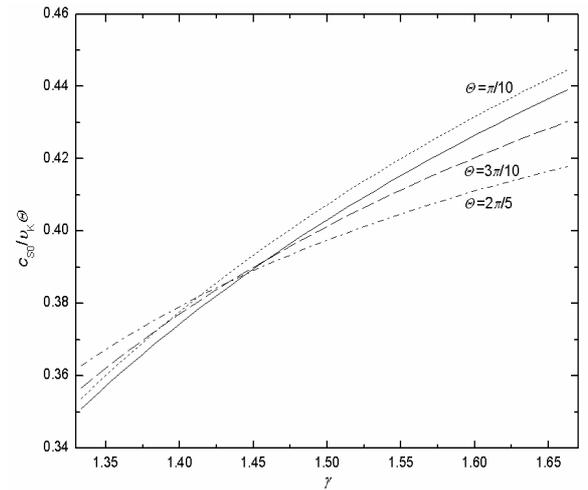

**Figure 2** The analytic relation $c_{s0}/\upsilon_K\Theta = [(\gamma-1)/2\gamma]^{1/2}$ (solid line) and the numerical variation of $c_{s0}/\upsilon_K\Theta$ with $\gamma$ for three different half-opening angles, $\Theta = \pi/10$ (dotted line), $\Theta = 3\pi/10$ (dashed line), and $\Theta = 2\pi/5$ (dot-dashed line).

In order to evaluate the validity of equation (12), we display the results in figure 2, which shows that the analytic relation of equation (12) (solid line) and the variation of $c_{s0}/\upsilon_K\Theta$ with $\gamma$ for three half-opening angles, $\Theta = \pi/10$ (dotted line), $\Theta = 3\pi/10$ (dashed line), and $\Theta = 2\pi/5$ (dot-dashed line). As shown in figure 2, the analytic relationship agrees well with the numerical results of equation (11) even for large half-opening angle $\Theta = 2\pi/5$.

To conclude, we revisit the vertical structure of accretion flows in spherical coordinates with considering the vertical distribution of $\upsilon_r$, $\upsilon_\varphi$, $c_s$, and $\rho$, and find that the analytic relation $c_{s0}/v_K \Theta = [(\gamma-1)/2\gamma]^{1/2}$ works well for both geometrically thin and thick accretion flows.